\documentclass[11pt, onecolumn]{IEEEtran}

\usepackage{psfrag}
\usepackage{cite}
\usepackage{url}
\usepackage{graphicx}
\usepackage{amsmath}
\usepackage{multirow}
\usepackage{array}
\usepackage{epsfig}
\usepackage{balance}
\usepackage{enumerate}
\usepackage{graphicx}
\usepackage{times}
\usepackage{amssymb}
\usepackage{algorithm}
\usepackage{algorithmic}
\usepackage{color}
\begin{document}
\title{Opportunistic Interference Alignment for Random Access Networks}
%\author{Author 1, Author 2, and Author 3}
\author{Hu Jin, {\em Member}, {\em IEEE}, Sang-Woon Jeon, {\em Member}, {\em IEEE}, and Bang Chul Jung, {\em Senior Member}, {\em IEEE}
\thanks{H. Jin is with the Department of Electronics and Communication Engineering, Hanyang University, Ansan, Republic of Korea (e-mail: hjin@hanyang.ac.kr).}
\thanks{S.-W. Jeon is with the Department of Information and Communication Engineering, Andong National University, Andong, Republic of Korea (e-mail: swjeon@anu.ac.kr).}%
\thanks{B. C. Jung (corresponding author) is with the Department of Information and Communication Engineering, Gyeongsang National University, Tongyeong, Republic of Korea (e-mail: bcjung@gnu.ac.kr).}}

%\IEEEpeerreviewmaketitle
\maketitle
\begin{abstract}
An interference management problem among multiple overlapped random access
networks~(RANs) is investigated, each of which operates with slotted ALOHA protocol.
Assuming that access points and users have multiple
antennas, a novel opportunistic interference alignment~(OIA) is proposed to mitigate
interference among overlapped RANs. The proposed technique intelligently combines the
transmit beamforming technique at the physical layer and the opportunistic packet
transmission at the medium access control layer. The transmit beamforming is based on
interference alignment and the opportunistic packet transmission is based on the
generating interference of users to other RANs, which can be regarded as a joint
optimization of the physical layer and the medium access control layer. It is shown
that the proposed OIA protocol significantly outperforms the conventional schemes such
as multi-packet reception and interference nulling.
\end{abstract}


\section{Introduction}
Recently, interference management in random access networks has
attracted a great interest. For an example,
the working group of IEEE 802.11,
which is one of the most successful standards in commercial wireless
communication systems, is considering performance improvement for
overlapping basic service sets (OBSS) under the new standardization
called IEEE 802.11 high efficiency wireless local area network
(HEW)~\cite{802.11hew}. It is known that OBSS experiences severe
interference and the interference among OBSS is a primary problem that
IEEE 802.11 HEW should overcome.

On the other hand, interference alignment (IA)~\cite{Jafar07} has been considered as a
promising solution to achieve the optimal degrees-of-freedom~(DoF) in several
interference network models including cellular networks. Suh and Tse characterized the
DoF of the $K$-cell uplink cellular network and they proposed an achievable scheme for
the optimal DoF~\cite{Suh08}. The main idea of \cite{Suh08} is to align the
interferences from users in other-cells to predefined interference spaces, and it was
shown that the achievable DoF increases as the number of users transmitting concurrently
increases. To apply IA, global channel state information
(CSI) is required and, hence, some performance degradation may occur as long as there
happens CSI feedback quantization in practice. The related performance was analyzed in
\cite{Chen14} along with a further performance optimization. In addition, opportunistic
IA~(OIA) has been proposed for the $K$-cell uplink network in which user scheduling is
combined with IA. Unlike the original IA technique, OIA does not require global CSI,
time/frequency expansion, and iterations for beamformer design, thereby resulting in
easier implementation~\cite{Jung11, Yang13}. Later,
\cite{Leithon12} compared OIA to the traditional IA with quantized CSI and indicated the
advance of OIA. To further improve the performance of OIA, an active alignment transmit
beamforming scheme was proposed in \cite{Gao13} which perfectly aligns the interference
to the reference interference direction of one BS and, therefore, achieves partial IA
with nonzero probability. As the previous OIA schemes mostly minimize the inter-cell
interference, a new OIA scheme that additionally considers the intra-cell power loss was
also  proposed in \cite{Ren14}.

%With the original IA technique, however, the global channel state information
%(CSI) is required at each transmitter in order to align interference signals at receivers.
%Moreover, the application of IA for
%the downlink cellular systems which is also known as interference broadcast channel (IBC) was also proposed in
%\cite{Suh11}.

IA techniques have been also applied to random access networks~(RANs) based on carrier
sensing multiple access (CSMA) mechanisms~\cite{Gollakota09, Lin11}. In
\cite{Gollakota09}, multiple packets from users in other overlapped networks are aligned
at the physical~(PHY) layer and the decoded packets are assumed to be exchanged through
wired backhaul such as Ethernet. Basically, at the medium access control~(MAC) layer in
\cite{Gollakota09}, users are scheduled like the point coordinated function (PCF) which
is a part of IEEE 802.11 standard. Thus, the proposed IA algorithm in \cite{Gollakota09}
requires tight coordination among access points~(APs) in overlapped networks through
wired backhaul and it does not consider collisions among users even though the collision
effect is the most important factor to degrade the performance of RANs. IA was applied
to a fully distributed random access environment in \cite{Lin11} like the distributed
coordinated function (DCF) of IEEE 802.11 standard. In \cite{Lin11}, the proposed MAC
protocol allows a new transmission even when there exist ongoing transmissions already
in overlapped networks as long as the new transmitter and receiver have a sufficiently
large number of antennas and ensure no interference to ongoing transmissions. Thus, if
users have the same number of antennas, then the performance improvement becomes
limited. Moreover, the DoF of multiple access channel was characterized when both users
and AP have multiple antennas and each user independently decides whether to transmit in
a specific time, which can be regarded as the uplink scenario of a single
RAN~\cite{Pourahmadi13}. In \cite{Pourahmadi13}, the authors proved the optimal average
DoF can be achieved by the interference alignment in specific network scenarios, but the
interference among multiple RANs is not considered.

%Interference management is one of the most crucial problems in wireless communication systems. Over the last
%decade, there has been a lot of effort on characterizing capacity of interference channels in terms of
%degree-of-freedom (DoF) which is also known as multiplexing gain.

In this paper, we propose a novel OIA protocol in order to efficiently manage the
interference among overlapped RANs operating in slotted ALOHA protocol. The proposed OIA
protocol jointly considers PHY and MAC layers. In the PHY layer, a beamforming algorithm
based on singular value decomposition (SVD) is adopted to minimize \textit{generating
interference} from each user to other overlapped RANs. In the MAC layer, a novel
opportunistic random access algorithm is proposed, which is based on cumulative
distribution function~(CDF) of each user's generating interference. Therefore, the
proposed OIA protocol is a cross-layer solution for interference-limited RANs.

The main differences between the proposed OIA protocol and the conventional OIA algorithm~\cite{Jung11, Yang13} are as follows.
\begin{itemize}
\item Each user determines to send packets for itself in the proposed OIA protocol, while base stations determine which users send packets in the conventional OIA algorithm.
\item The proposed OIA protocol utilizes the CDF value of each user's generating interference as random access criterion, while the conventional OIA algorithm utilizes the very generating interference as a scheduling criterion.
\item The number of concurrently transmitting users in the network is a random variable in the proposed OIA protocol, while it is fixed in the conventional OIA algorithm. Thus, the proposed OIA protocol examines the number of concurrently transmitting users in the network and changes the packet decoding methodology according to the number of concurrently transmitting users.
\end{itemize}

The rest of this paper is organized as follows:
Section~\ref{sec:System_model} describes the system model and
Section~\ref{sec:Conv} introduces the conventional techniques which can
be used for interference-limited RANs. In Section~\ref{sec:OIA}, the
OIA protocol for the interference-limited RANs is proposed.
Section~\ref{sec:Evaluation} evaluates the throughput performance of
the proposed OIA protocol and, finally, Section~\ref{sec:Conclusion}
concludes the paper.

\begin{figure}
\centering
\includegraphics[width= 4in]{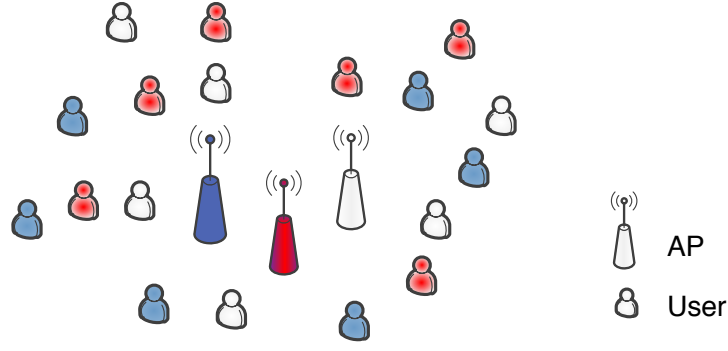}
    \caption{Illustration of overlapped RANs~($K=3$).}
    \label{fig:Scenario}
\end{figure}

\section{System model}\label{sec:System_model}
We consider an \textit{uplink} scenario in $K$ overlapped RANs each of
which has one AP and $N$ users. Each AP and each user are assumed to
have $M$ and $L$ antennas, respectively. Fig.~\ref{fig:Scenario} shows
an example network where three RANs are overlapped. We assume that each
RAN operates with slotted ALOHA and transmission time is equally
divided by slots. At each slot, each user transmits a packet to its
serving AP with probability $p$. Due to the nature of random access,
simultaneous transmissions from multiple users is inevitable. Assuming
that RANs are geometrically overlapped, such concurrent packet
transmissions from multiple users in different RANs cause interference
due to the broadcasting property of wireless medium. The channel matrix
from the $j$-th user in the $i$-th RAN to the $k$-th AP is denoted by
$\textbf{H}^{[i,j]}_k \in \mathbb{C}^{M\times L}$, where $i, k \in \{1,
. . .,K\}$ and $j \in \{1, . . .,N\}$. The received signal at the
$k$-th AP, $\textbf{y}_{k} \in \mathbb{C}^{M\times 1}$, is given as
\begin{equation}
\textbf{y}_{k} = \sum_{i=1}^K \sum_{j=1}^N \textbf{H}^{[i,j]}_k
\textbf{w}^{[i,j]} x^{[i,j]} s^{[i,j]}+\textbf{z}_k,
\end{equation}
where $s^{[i,j]} \in \{0,1\}$ denotes the random variable representing the activity of
the $j$-th user in the $i$-th RAN and $\mathsf{Pr}\{s^{[i,j]}=1\}=p$. Hence, $s^{[i,j]}$
becomes zero if the user has no packet to transmit. $x^{[i,j]} \in \mathbb{C}$ and
$\textbf{w}^{[i,j]} \in \mathbb{C}^{L\times 1}$ denote the information stream and its
transmit beamforming vector of the $j$-th user in the $i$-th RAN, respectively. Let
$s_i\triangleq \sum_{j=1}^{N} s^{[i,j]}$ and $s \triangleq \sum_{i=1}^{K} s_i$. That is,
$s_i$ indicates the number of concurrently transmitting users in the $i$-th RAN and $s$
indicates the total number of concurrently transmitting users in all $K$ RANs. We assume
that each user experiences Rayleigh fading and the channel gain independently changes in
each time slot. Each AP is assumed to periodically transmit a pilot signal so that the
users can estimate their channels to all APs located in the overlapped area by using the
reciprocity of the wireless channel. In this paper, we also assume that each user
transmits a single information stream. $\textbf{z}_k\in \mathbb{C}^{M\times 1}$
represents the additive Gaussian noise at the $k$-th AP. Furthermore, the $j$-th user in
the $i$-th RAN is assumed to know its outgoing channels $\textbf{H}^{[i,j]}_k$, $k \in
\{1,2,...,K\}$, by exploiting the channel reciprocity, i.e., the \textit{local CSI} at
the transmitter. In addition, we assume the interference-limited networks and the
average path-loss from users to APs are assumed to be identical to each other for
simplicity.

\section{Conventional Techniques for Interference-Limited RANs}\label{sec:Conv}
\subsection{Multi-Packet Reception~(MPR)}\label{sec:MPR}
For designing RANs, the mitigation of collision effects among simultaneously
transmitting users is one of the most challenging issues. It has been shown that
multi-packet reception~(MPR) at the PHY layer, which is implemented through multi-user
multiple-input multiple-output (MIMO) techniques in general, can be a promising
solution~ \cite{Jin08}. As assumed in Section~\ref{sec:System_model}, each user
transmits a single stream and each AP has $M$ antennas, each AP can successfully decode
up to $M$ packets from transmitting users in the overlapped area by using the MPR
technique~\cite{Jin11}. In practice, the probability that each AP successfully decodes
the received packets depends on the number of concurrently transmitting users. The
average throughput (packets/slot) in the MAC layer of $K$-overlapped RANs with the MPR
technique is obtained as
\begin{equation}
T_{\rm MPR} = \sum_{m=1}^M m \cdot {NK\choose m} p^m (1-p)^{NK - m}
\cdot P_{m, M}^{\rm MPR},
\end{equation}
where $NK$ denotes the total number of users in the $K$-overlapped RANs and $P_{m,
M}^{\rm MPR}$ denotes the probability that each received packet is successfully decoded
when there exist $m$ concurrently transmitting users in the whole $K$-overlapped RANs.
Note that if the $k$-th RAN has $m_k$ users'
simultaneous transmissions, $m = \sum_{k=1}^K m_k$. Although the $k$-th AP may try to
decode all the $m$ users' packets, only the $m_k$ packets are useful for it and it may
throw the other decoded packets. Obviously, $P_{m, M}^{\rm MPR}$ decreases as $m$
increases due to the interferences from $m-1$ users. More detailed analysis on $P_{m,
M}^{\rm MPR}$ is given in \cite{Jin08, Jin11}.

%\begin{equation}
%T_{\rm MPR, ideal} = \sum_{m=1}^M m \cdot {NK\choose m}  p^m (1-p)^{NK - m},
%\end{equation}
%where $NK$ denotes the total number of users in the $K$ overlapping RANs.
%
% As the number increases, the success probability surely decreases due to the interference among users. Let the success probability $P_{m, M}^{\rm MPR}$ when there exist $m$ concurrently transmitting users in the network, the throughput can be expressed as
%\begin{equation}
%T_{\rm MPR} = \sum_{m=1}^M m \cdot {NK\choose m} p^m (1-p)^{NK - m} \cdot P_{m, M}^{\rm MPR},
%\end{equation}
%where more detailed analysis on $P_{m, M}^{MPR}$ is given in \cite{Jin08, Jin11}.

\begin{figure}
\centering
\includegraphics[width= 4.0in]{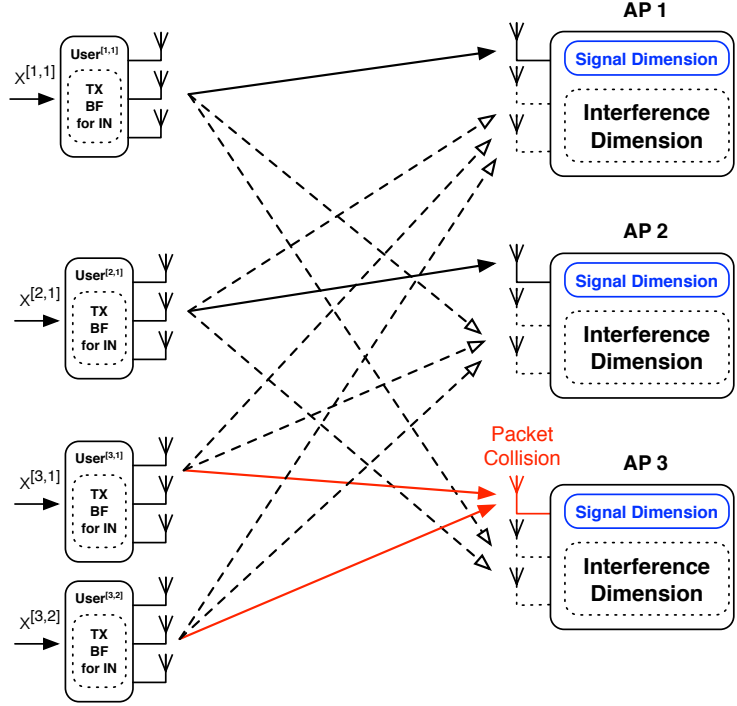}
    \caption{Transmission scenario with the IN ($K=3$, $L=3$, $M=3$).}
    \label{fig:IN_model}
\end{figure}

\subsection{Interference Nulling~(IN)}\label{sec:IN}
%It is easy to observe that the MPR scheme does not fully exploit the potential of the multiple antennas at each
%user.
In addition to MPR, we can also consider the interference nulling~(IN)
technique which utilizes the transmit beamforming at each user for
overlapped RANs. For applying the IN technique, each AP sets $S$
spatial dimensions for receiving signals from its serving users.
%Without loss of generality, the $m$-th antenna at the AP is regarded as the $m$-th spatial dimension in this paper.
Then, each user performs transmit beamforming to make its signal
arriving at the signal spaces of the other APs be zero. Assuming $K$
overlapped RANs and $S$ dimensional signal space at each AP, each user
should make its transmit signal, which is arrived at $(K-1)S$
dimensional signal space in other APs, be zero. Hence, the dimension of
signal space at each AP, $S$, should satisfy the following constraint:
\begin{equation} \label{eq:IN_limitation}
S < \min\left\{\frac{L}{K-1},M \right\}
\end{equation}
where $L$ indicates the number of transmit antennas at users.
Fig.~\ref{fig:IN_model} shows a transmission scenario with the IN
technique when $K=3$, $L=3$, and $S=1$, which satisfies the condition
in (\ref{eq:IN_limitation}). In this figure, it is assumed that
$s_1=1$, $s_2=1$, and $s_3=2$. In Fig.~\ref{fig:IN_model}, without loss
of generality, each AP is assumed to set its first antenna as the
signal space for receiving packets from its serving users, while other
two antennas are reserved for interference signals from other RANs. All
transmitting users in the overlapped networks perform transmit
beamforming in order to null out the interference at other APs' signal
spaces, which is the first antenna of each AP in
Fig~\ref{fig:IN_model}. Each AP receives the signals from the users
belong to itself through the first antenna, and each AP operates as an
independent cell since there is no interference from other cells for
the first antenna. In Fig.~\ref{fig:IN_model}, there is a
\textit{packet collision} in the third RAN, but the APs in the first
and second RANs can receive a packet successfully since interference
signals are still nulled out at the signal spaces.
%As a representative example, the transmission behavior at the user 3 is depicted in Fig.~\ref{fig:IN_model}.

\begin{figure}[t]
\centering
\includegraphics[width= 3.0in]{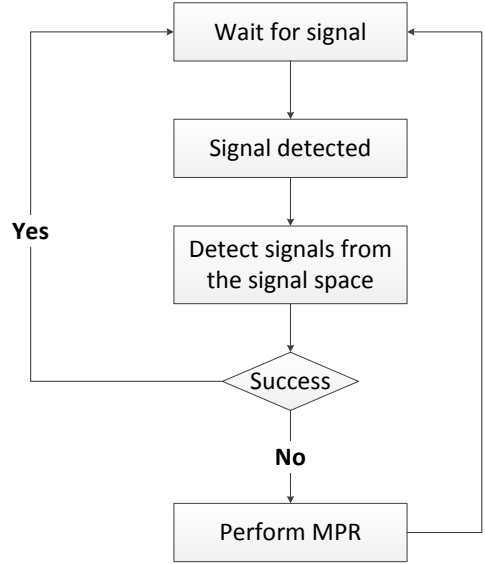}
    \caption{Signal detection procedure at each AP with the IN technique.}
    \label{fig:IN_detection}
\end{figure}

If the number of concurrently transmitting users in a specific RAN is
less than or equal to $S$, that is, $s_i \le S $, $i \in
\{1,2,...,K\}$, then each AP can surely decode the received signals
with the MPR technique \textit{regardless of the number of transmitting
users in other RANs}. However, if the number of concurrently
transmitting users in a specific RAN is larger than $S$ but the number
of concurrently transmitting users in all the $K$ RANs is smaller than
$M$, that is, $s_i > S$ but $s \le M$, $i \in \{1,2,...,K\}$, then the
AP can still successfully decode the received signal by using all
receive antennas with MPR.
%If the total number of transmitting users in the overlapping cells is smaller
%than $M$, the BS still has the possibility to detect the signals in its cell with MPR.
%Reflecting this phenomenon,
The overall signal detection procedure of the IN technique is shown in
Fig.~\ref{fig:IN_detection}.

%
%Assuming a perfect physical layer as in Section \ref{sec:MPR}, the MAC layer throughput is obtained as
%\begin{equation} \label{eq:IN_perfect}
%\begin{array}{ll}
%T_{\rm IN, ideal}  &\displaystyle = K \cdot \left\{ \sum\limits_{m = 1}^S {m {N\choose m} p^m \left( {1 - p} \right)^{N - m} }  \right. \\
%        &\displaystyle + \sum\limits_{m = S+1}^M m \cdot {N\choose m}p^m \left( {1 - p} \right)^{N - m} \cdot \\
%        &\displaystyle \left. \left[ \sum\limits_{j = 0}^{M - m} {{N(K-1) \choose j}p^j (1 - p)^{N ( K - 1) - j} }
%\right]  \right\}.
%\end{array}
%\end{equation}
%In the curly bracket of (\ref{eq:IN_perfect}), the first summation shows the throughput obtained from the signal
%space and the throughput in the second summation is obtained from the additional MPR procedure with all $M$
%antennas.
By taking into account this possibility, the throughput at the MAC
layer with the IN technique is given as
\begin{equation} \label{eq:IN_imperfect}
\begin{array}{ll}
 T_{\rm IN}  =& K \cdot \left\{ \sum\limits_{m = 1}^S {m \cdot {N\choose m} p^m \left( {1 - p} \right)^{N - m} \cdot P_{m, S}^{\rm MPR}}  \right. \\
        &  \displaystyle + \sum\limits_{m = S+1}^M m \cdot {N\choose m}p^m \left( {1 - p} \right)^{N - m} \cdot  \displaystyle \left. \left[ \sum\limits_{j = 0}^{M - m} {{N(K-1) \choose j}p^j (1 - p)^{N ( K - 1) - j} \cdot  P_{m+j, M}^{\rm MPR}}
\right]  \right\},
\end{array}
\end{equation}
where $P_{m, S}^{\rm MPR}$ denotes the probability that the received signals are
successfully decoded when $m$ and $j$ users concurrently
transmit packets in a specific RAN and the other RANs respectively and each AP sets $S$
antennas as the signal space. Here, $S$ is assumed to satisfy the condition in
(\ref{eq:IN_limitation}). In \eqref{eq:IN_imperfect}, the
first summation in the brace shows the throughput of each RAN obtained from the signals
arriving at the $S$-dimensional signal space. The second summation shows the throughput
obtained when the number of simultaneously transmitting users in a RAN is larger than
$S$ but the total number of simultaneously transmitting users in the whole $K$ RANs,
$s$, is less than $M$. Note that the AP may possibly decode those packets by performing
MPR with its whole $M$ antennas if the concurrently transmitting users in the $K$ RANs
are no larger than $M$. In \eqref{eq:IN_imperfect}, $m\cdot{N \choose m}p^m (1-p)^{N-m}$
is the probability that $m$ users are simultaneously transmitting in a RAN and the
formula in the bracket shows the probability that the number of simultaneously
transmitting users in other $K-1$ RANs is no larger than $M-m$. As we consider the total
throughput of the $K$-overlapping RANs, the factor of $K$ is appeared outside of the
brace in \eqref{eq:IN_imperfect}.

\section{Opportunistic Interference Alignment}\label{sec:OIA}
As mentioned before, the constraint on the dimension of signal space, which is shown in
(\ref{eq:IN_limitation}), limits the DoF in each RAN and restricts the applicability of
the IN technique in practice. In this section, we propose the OIA protocol which
efficiently controls the interference among overlapped RANs. In the proposed OIA
protocol, the DoF in each RAN, $S$, is not limited by (\ref{eq:IN_limitation}) and can
be arbitrarily set from $1$ to $M$. The OIA protocol is designed by considering both the
PHY and MAC layers, which is as follows.
%To overcome this drawback, in this section we propose an opportunistic interference
%alignment (OIA) scheme which consists of two pars: the SVD-based Beamforming at the physical layer shown in
%Section~\ref{subsec:SVD_beamforming} and the opportunistic transmission mechanism at the MAC layer shown in
%Section~\ref{subsec:Opportunism}.

\subsection{PHY Layer: SVD-Based Transmit Beamforming}~\label{subsec:SVD_beamforming}
First of all, the $k$-th AP sets its interference space for
interference alignment, which is denoted by
$\mathbf{Q}_k=[\mathbf{q}_{k,1}, ...,\mathbf{q}_{k,M-S}]$, where
$\mathbf{q}_{k,m} \in \mathbb{C}^{M \times 1}$ is the orthonormal
basis, $1 \le k \le K$ and  $1 \le m \le M-S$. Here, $S \in
\{1,2,...,M\}$ denotes the dimension of the signal space reserved at
each AP and it is assumed that $s_i=S$ for all $i\in\{1,2,...,K\}$ for
convenience in this subsection since we focus on the operation at the
PHY layer. Obviously, $s_i$ can vary from $0$ to $N$ according to the
MAC layer operation which is the focus of the next subsection. %Hence, we consider the PHY layer operation for given $S$ active users in each RAN.
%Each AP independently generates $\mathbf{q}_{k,m}$ from the isotropic distribution over the $M$-dimensional unit sphere.
For given $\mathbf{Q}_{k}$, the $k$-th AP
also calculates the null space of $\mathbf{Q}_{k}$, defined by
 \begin{equation}
 \mathbf{U}_k = \left[ \mathbf{u}_{k,1}, \ldots, \mathbf{u}_{k,S} \right] \triangleq \textrm{null}(\mathbf{Q}_k),
 \end{equation}
 where $\mathbf{u}_{k,i}\in \mathbb{C}^{M \times 1}$ is the orthonormal basis, and broadcasts it to all users in the network.\footnote{By using the pre-defined pseudo-random pattern, $\mathbf{Q}_{k}$ can be informed to users without any signaling process and each user may compute $\mathbf{U}_k$ in a distributed manner without being informed.}
If $S=M$, then $\mathbf{U}_k$ can be any orthonormal matrix.

%Each AP reserves $S~(1 \le S \le M)$ spatial spaces for receiving the signals from users belong to itself and broadcasts the .
%Note that there is no constraint for $S$.
We assume the unit-norm beamforming vector at the $j$-th user in the
$i$-th RAN as $\mathbf{w}^{[i,j]}$, i.e., $\left\| \mathbf{w}^{[i,j]}
\right\|^2 = 1$. From $\mathbf{U}_k$ and $\mathbf{H}^{[i,j]}_{k}$, the
$j$-th user in the $i$-th RAN calculates its \textit{effective}
generating interference, called \textit{leakage of interference~(LIF)},
from
\begin{align}
\eta^{[i,j]}_{k} &= \left\|\textrm{Proj}_{\bot \mathbf{Q}_k}\left( \mathbf{H}_{k}^{[i,j]}\mathbf{w}^{[i,j]}\right)\right\|^2 \nonumber \\
%& = \sum_{m=1}^{S}\left\| \left({\mathbf{u}_{k,m}}^{H}\mathbf{H}_{k}^{[i,j]} \mathbf{w}^{[i,j]} \right){\mathbf{u}_{k,m}} \right\|^2 \\
\label{eq:eta_tilde}&= \left\|\mathbf{U}_k^{H}\mathbf{H}_{k}^{[i,j]}
\mathbf{w}^{[i,j]} \right\|^2,
\end{align}
where $i\in \{1,2,...,K\}$, $j \in \{1,2,...,N\}$, and $k\in
\{1,2,...,K\}\setminus i= \{1, \ldots, i-1, i+1, \ldots, K\}$. Here,
$\textrm{Proj}_{\bot \mathbf{A}}(\mathbf{B})$ denotes the projection
operation of $\mathbf{B}$ onto the null space of $\mathbf{A}$ and
$(\cdot)^H$ denotes the Hermitian operation. LIF can be regarded as the
interference power which is received at the $k$-th AP and not aligned
at the interference space $\mathbf{Q}_k$.

%If $S<M$ and the interference from the $j$-th user in the $i$-th RAN to
%the $k$-th AP is perfectly aligned to $\mathbf{Q}_k$, i.e.,
%\begin{equation} \label{eq:IA_span}
%\mathbf{H}^{[i,j]}_{k}\mathbf{w}^{[i,j]}\in \textrm{span} \left[
%\mathbf{Q}_k \right],
%\end{equation}
% then it is nulled in the signal space of the $k$-th AP because ${\mathbf{U}_k}^{H} \mathbf{H}^{[i,j]}_{k}\mathbf{w}^{[i,j]} = \mathbf{0}$, i.e., ${\eta}^{[i,j]}_{k} = 0$.   If $S=M$, the LIF is simplified to ${\eta}^{[i,j]}_{k}= \left\|\mathbf{H}_{k}^{[i,j]} \mathbf{w}^{[i,j]} \right\|^2$.

Instead of perfectly nulling interference from users to other RANs, the
proposed transmit beamforming is performed to minimize sum of the
effective interference power to other APs. Thus, each user finds the
optimal transmit beamforming vector $\mathbf{w}^{[i,j]}$ that minimizes
its LIF metric which defined as:
\begin{align}\label{eq:LIF}
\eta^{[i,j]}_{\textrm{\textrm{SUM}}} &= \sum_{k=1, k\neq i}^{K} \left\|
{\mathbf{U}_k}^{H}\mathbf{H}_{k}^{[i,j]}\mathbf{w}^{[i,j]}\right\|^2
\triangleq \left\| \mathbf{G}^{[i,j]} \mathbf{w}^{[i,j]}\right\|^2,
\end{align}
where $\mathbf{G}^{[i,j]}\in \mathbb{C}^{(K-1)S\times L}$ is defined by
\begin{align} \label{eq:G_def}
\mathbf{G}^{[i,j]} &\triangleq \Bigg[ \left({\mathbf{U}_1}^{H}\mathbf{H}_{1}^{[i,j]}\right)^{T}, \ldots, \left({\mathbf{U}_{i-1}}^{H}\mathbf{H}_{i-1}^{[i,j]}\right)^{T},  \nonumber\\
&
\hspace{20pt}\left({\mathbf{U}_{i+1}}^{H}\mathbf{H}_{i+1}^{[i,j]}\right)^{T},
\ldots, \left({\mathbf{U}_K}^{H}\mathbf{H}_{K}^{[i,j]}\right)^{T}
\Bigg]^{T}.
\end{align}

Let us denote the SVD of $\mathbf{G}^{[i,j]}$ as
\begin{equation} \label{eq:G_SVD}
\mathbf{G}^{[i,j]} =
\boldsymbol{\Omega}^{[i,j]}\boldsymbol{\Sigma}^{[i,j]}{\mathbf{V}^{[i,j]}}^{H},
\displaybreak[0]
\end{equation}
where $\boldsymbol{\Omega}^{[i,j]}\in \mathbb{C}^{(K-1)S\times L}$ and
$\mathbf{V}^{[i,j]}\in \mathbb{C}^{L\times L}$ consist of $L$
orthonormal columns, respectively, and $\boldsymbol{\Sigma}^{[i,j]} =
\textrm{diag}\left( \sigma^{[i,j]}_{1}, \ldots,
\sigma^{[i,j]}_{L}\right)$, where $\sigma^{[i,j]}_{1}\ge \cdots
\ge\sigma^{[i,j]}_{L}$. \pagebreak[0] Then, it is apparent that the
optimal $\mathbf{w}^{[i,j]}$ is determined as
\begin{equation} \label{eq:W_SVD}
\mathbf{w}^{[i,j]}_{\textrm{SVD}} = \arg  \min_{\mathbf{w}^{[i,j]}}
\left\| \mathbf{G}^{[i,j]} \mathbf{w}^{[i,j]}\right\|^2
=\mathbf{v}^{[i,j]}_{L},
\end{equation}
where $\mathbf{v}^{[i,j]}_{L}$ is the $L$-th column of
$\mathbf{V}^{[i,j]}$. With this choice, $\eta^{[i,j]}_{\textrm{SUM}} =
{\sigma^{[i,j]}_{L}}^2$ is achievable.
%\begin{equation} \label{eq:LIF_beamforming_simple}
%\eta^{[i,j]}_{\textrm{SUM}} = {\sigma^{[i,j]}_{L}}^2.
%\end{equation}

% The $j$-th user in cell $i$ who tries to transmit signals to its
%serving BS, the SVD-based beamforming is performed as follows: First, it constructs interference matrix $G^{[i,
%j]}$ to other BSs as follows:
%\begin{equation}
%G^{[ {i,j} ]}  = \left[ \left( {H_1^{[ {i,j} ]} } \right)^T , \cdots ,\left( {H_{i - 1}^{[ {i,j} ]} } \right)^T
%,\left( {H_{i + 1}^{[ {i,j} ]} } \right)^T , \cdots ,\left( {H_K^{[ {i,j} ]} } \right)^T \right]^T
%\end{equation}
%Then, it performs SVD for the interference matrix,
%\begin{equation}
%G^{\left[ {i,j} \right]}  = \Omega ^{[i,j]} \Sigma ^{[i,j]} V^{\left[ {i,j} \right]}. _{\,\,}
%\end{equation}
%where $ \Omega ^{[i,j]}$ and $ V ^{[i,j]}$ are unitary matrices and $\Sigma ^{[i,j]}$ is a diagonal matrix whose
%elements in the diagonal are the singular values.
%
%In order to minimize the interference to other cells, the precoding vector is chosen to the vector corresponds to
%the minimum singular value, i.e.,
%\begin{equation}
%w_{\rm SVD}^{\left[ {i,j} \right]}  = \arg \mathop {\min }\limits_v \left\| {G^{\left[ {i,j} \right]} v} \right\|^2
%= v_L^{\left[ {i,j} \right]}.
%\end{equation}

After receiving signals at the $k$-th AP, $\mathbf{U}_k$ is multiplied
to remove the interference that is aligned to the interference space of
the $k$ AP, $\mathbf{Q}_k$. Then, the received signal can be expressed
as:
\begin{eqnarray} \label{eq_y2}
\mathbf{\tilde{y}}_{k} &=& \mathbf{U}_k \mathbf{y}_{k}.
%&=&  \sum_{i=1}^K \sum_{j=1}^N \textbf{H}^{[i,j]}_k \textbf{w}^{[i,j]} x^{[i,j]} s^{[i,j]}+\textbf{z}_k,
\end{eqnarray}

\begin{figure*}
\centering
\includegraphics[width=16.5cm]{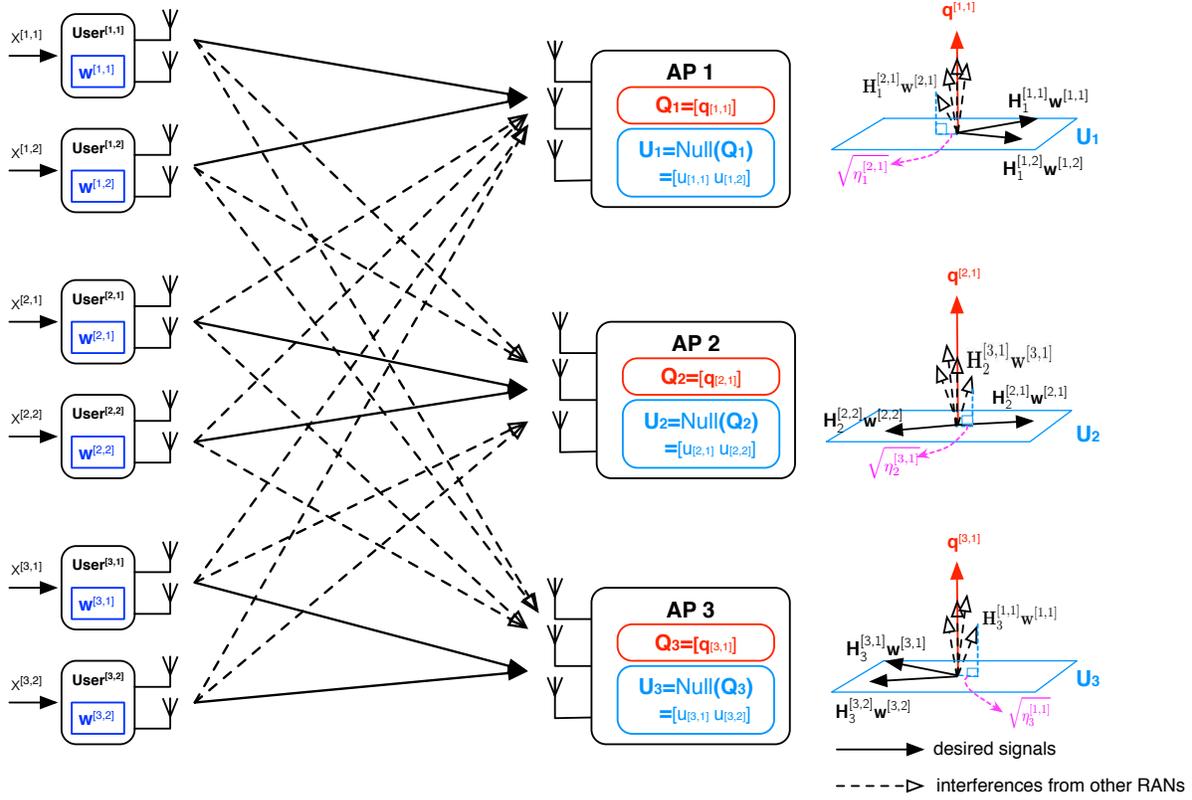}
    \caption{Geometric interpretation of the proposed opportunistic interference alignment technique ($K=3$, $L=3$, $M=3$, $N=2$, $S=2$).}
    \label{fig:OIA_model}
\end{figure*}

When $S < \frac{L}{K-1}$, this beamforming makes the interference
signals be zero. Thus, the proposed OIA includes the IN technique
presented in Section~\ref{sec:IN}. On the other hand, when $S \ge
\frac{L}{K-1}$, the users' transmissions may cause interference to the
packet received at other APs. Fig.~\ref{fig:OIA_model} shows the
geometric signal structure of the proposed OIA protocol at APs when
$K=3$, $L=M=3$, $N=2$ and $S=2$, which corresponds to the case where
$S\geq \frac{L}{K-1}$ and, as a result, the IN technique cannot be applied. For simplicity, we assume that the interference space of each
AP is the same in Fig.~\ref{fig:OIA_model}. Note that the proposed
transmit beamforming minimizes the LIF metric in \eqref{eq:LIF}.
%The BSs can perform MIMO decoding techniques such as zero forcing to decode packets.
In the next subsection, we will introduce an opportunistic random
access mechanism to further reduce such interference.

\subsection{MAC Layer: Interference-Aware Opportunistic Transmission}~\label{subsec:Opportunism}
Although the SVD-based transmit beamforming minimizes the interference
to other RANs at the PHY layer, the residual interference may exist in
the signal space at APs. Hence, we need further reduce the interference
at the MAC layer by exploiting  opportunistic random access. In the
conventional OIA technique proposed for cellular networks, the number
of transmitting users in each cell is fixed according to the
predetermined scheduling policy. In RANs, however, the number of
concurrently transmitting users in each RAN is a random variable which
can vary over time by nature. While the conventional opportunistic
random access in RANs maximizes the signal strength of
users~\cite{Hwang09}, each user applies the opportunism based on its
effective generating interference to other RANs, i.e., the  LIF metric
in (\ref{eq:LIF}), in the proposed OIA protocol.

%We define the amount of LIF as follows:
%\begin{equation}
%\eta ^{\left[ {i,j} \right]}  = \sum\limits_{k = 1,k \ne i}^K {\left\| {H_k^{\left[ {i,j} \right]} w_{\rm
%SVD}^{\left[ {i,j} \right]} } \right\|^2 }.
%\end{equation}

Each user observes its LIF metric for a long time to obtain the corresponding cumulative
distribution function (CDF). Specifically, in each time
slot, each user calculates its instant LIF metric based on (\ref{eq:LIF}) and stores it
to update the histogram of the LIF values from which the CDF can be calculated by
normalization. Another possible methodology to obtain the CDF is as follows: Let
$F_{t-1}(\eta)$ be the stored CDF at slot $t-1$ and the LIF calculated at slot $t$ be
$\eta_0$.  Then the CDF can be updated as
\begin{equation}
F_{t}(\eta) = \left\{
\begin{array}{ll}
\displaystyle \frac{W F_{t-1}(\eta)}{W+1}, & \eta < \eta_0, \\
\displaystyle \frac{W F_{t-1}(\eta) + 1}{W+1}, & \eta\ge \eta_0,
\end{array}
 \right.
\end{equation}
where $W$ is the observation window. With larger $t$ and $W$, we can
obtain a more accurate CDF.

Once the CDF is determined, each user finds the output of the obtained CDF by using the
instant LIF metric. In the proposed OIA protocol, each user transmits a packet if the
output of the CDF is smaller than a certain threshold. Note that the CDF values are
uniformly distributed in $[0, 1]$~\cite{Jin14}. The transmission probability of each
user becomes $p$ by setting the threshold to $p$. If a
user already has a steady state CDF of $F(\eta)$, the equivalent operation of the
proposed protocol is that the user transmits its packet if the current LIF metric is
smaller than $F^{-1}(p)$. As long as the number of simultaneously transmitting users in
a RAN is smaller than $S$, the AP may decode the desired streams by treating the signals
from other cells as interference which is minimized by the opportunistic transmission at
the MAC layer and the SVD-based beamforming at the PHY layer. 

As the number of users in each RAN, $N$, increases, the transmission probability, $p$,
should be decreased in order to avoid packet collisions among users. Then, the decrease
of $p$ also leads to the decrease of the generating interference
of each user to other RANs. Hence, we can conclude that
the interference among RANs would be reduced as $N$ increases since a smaller $p$ makes
the LIF become smaller.

In the proposed OIA protocol, each user needs to calculate its instant LIF
value and find the corresponding CDF value in each time slot. This process may result in
computational burden to users since channel estimation and SVD operation are required.
However, most interference management techniques, including the conventional OIA
algorithms proposed for cellular uplink/downlink networks, require users to estimate the
interference channels to other cells for exploiting opportunistic user scheduling. In
addition, obtaining CDF value does not yield severe computational burden at each user,
compare to the channel estimation operation. Thus, the proposed OIA protocol has a
similar computational complexity with the conventional OIA algorithms. Note that the
proposed OIA does not require users to feed back channel matrix or scheduling metric,
i.e., CDF value, to their APs, while the conventional OIA algorithms in cellular
networks require all users to feed back their LIF and beamforming vector to the
corresponding base stations. Therefore, the proposed OIA protocol does not increase
computational complexity much at each user, and it can be applied to practical RANs
(such as wireless LANs) without significant modifications.

\subsection{Throughput Analysis}~\label{subsec:Throughput_OIA}
If the number of concurrently transmitting users in each RAN is smaller
than $S$, $1\leq S\leq M$, and the residual interference at APs is
small enough, then the throughput of the OIA  protocol can be expressed
as a similar form in (4).
%If we assume that the residual interference at APs in the proposed OIA protocol is small enough so that the BSs can successfully
%decode the transmitted packets in its RAN when the number of transmitting users in each RAN are smaller than $S$, the throughput of the OIA protocol can be expressed as (\ref{eq:IN_imperfect}) for a given $S$, $1 \le S \le M$.
Note that the constraint shown in (\ref{eq:IN_limitation}) does not
limit $S$ in the proposed OIA protocol. In practice, however, the
residual interference at the APs may reduce the packet success
probability. The packet success probability of the $i$-th AP may depend
on the dimension of reserved signal space~($S$), the number of
concurrently transmitting users in the $i$-th RAN~($s_i$), the number
of receive antennas at AP~($M$), the number of transmit antennas at
each user~($L$), and the number of concurrently transmitting users in
other RANs~($s-s_i$). Hence, the throughput of the proposed OIA
protocol can be expressed as:
\begin{equation} \label{eq:T_OIA}
\begin{array}{ll}
T_{\rm OIA}
\!\!\!\!\!\!&\displaystyle = K \cdot \left\{ \sum\limits_{m = 1}^M m \cdot {N\choose m}p^m \left( {1 - p} \right)^{N - m} \cdot \right.\\
 &\displaystyle  \!\!\!\!\!\!\!\!\!\!\!\!\! \left[ \sum\limits_{j = 0}^{M - m} {{N(K-1) \choose j}p^j (1 - p)^{N ( K - 1) - j} \cdot  P^{\rm MPR}_{m+j, M}} \right] \\
        &\displaystyle  + \sum\limits_{m = 1}^S m \cdot {N\choose m}p^m \left( {1 - p} \right)^{N - m} \cdot\\
        &\displaystyle \!\!\!\!\!\!\!\!\!\!\!\!\! \left. \left[ \sum\limits_{j = M-m + 1}^{N(K-1)} \!\!\!\! {{N(K-1) \choose j}p^j (1 - p)^{N ( K - 1) - j} \cdot  P^{\rm OIA}_{m, j}} \right] \right\},
\end{array}
\end{equation}
where $P^{\rm OIA}_{m, j}$ denotes the probability that the received packets are
successfully decoded, when there exist $m$ concurrently transmitting users in a specific
RAN and $j$ concurrently transmitting users exist in other RANs for given $L$, $S$, and
$M$. If the total concurrently transmitting users in whole networks~($s=m+j$) is less
than the number of receive antennas at APs~($M$), then the MPR technique can be used for
decoding the received packets. This phenomenon results in
the throughput of a single RAN shown by the first summation in the brace of
(\ref{eq:T_OIA}). On the other hand, if $s>M$ and $s_i \le S$, then the probability
that the received packets are successfully decoded is determined by the proposed OIA
protocol which includes the transmit beamforming, opportunistic access, and receive
beamforming. The corresponding throughput is described
by the second summation in the brace of (\ref{eq:T_OIA}). Note that here only the
scenarios where $s=m+j \ge M+1$ are reflected to exclude the cases already considered in
the first summation. Unfortunately, $P^{\rm OIA}_{m,j}$ is not mathematically tractable
due to complicated interactions among many factors such as the time-varying nature of
the number of active users in RANs, and it should be evaluated by simulations. In
Section~\ref{sec:Evaluation}, we evaluate $P^{\rm OIA}_{m,j}$ by simulation and
demonstrate the resultant throughput of the proposed OIA protocol in various
environments.
%In Section~\ref{sec:Evaluation}, we illustrate the resultant throughput of the proposed OIA protocol in various environments.
%Note that $m+j$ is larger than $M$ when the probability$P^{\rm OIA}_{m, j, M}$ is applied.
%Moreover, the effect of the
%opportunistic random access in MAC layer and the the SVD-based beamforming performed by the $j$ users should be considered in
%evaluating $P^{\rm OIA}_{m, j, M}$.
%As the number of transmitting users in the other networks varies over time in
%random access networks, the values of $P^{\rm OIA}_{m, j, M}$ cannot be analytically analyzed and, in this paper,
%their effect is considered in the simulations which is performed to obtain the throughput.
%In short, there exist two main contributions for obtaining the throughput of the proposed OIA protocol, which is shown in \eqref{eq:T_OIA}: one is MPR decoding and the second is from the ZF decoding performed at the signal space where the interference from other networks is minimized with the
%SVD-based beamforming.

\section{Numerical Results}~\label{sec:Evaluation}
We first consider a three-overlapped network each of which supports
$10$ users~($K=3$ and $N=10$). Both APs and users have three
antennas~($M=L=3$) and the channel matrix between each
transmit--receive antenna pair is assumed to experience Rayleigh
fading. The average received signal-to-noise ratios (SNRs) at APs from
all users in the network are assumed to be $0$dB. The APs adopt
zero-forcing~(ZF) technique for decoding the multiple received signals
in all protocols including MPR, IN, and the proposed OIA.
In particular, in the proposed OIA protocol, the ZF decoder is used for decoding packets from users in the corresponding RAN after null projection $\mathbf{U}_k$ as shown in \eqref{eq_y2}.
The
signal-to-interference-plus-noise ratio (SINR) threshold for successful
packet decoding is assumed to be $0$dB.\footnote{This value can be
varied according to PHY layer data-rate at transmitters, but we use a
single threshold in this paper since we focus on the MAC layer
throughput. Furthermore, we found that the performance tendencies in
this section are not changed according to various threshold through
extensive simulations.}

\begin{figure}
\centering
\includegraphics[width=4.5 in]{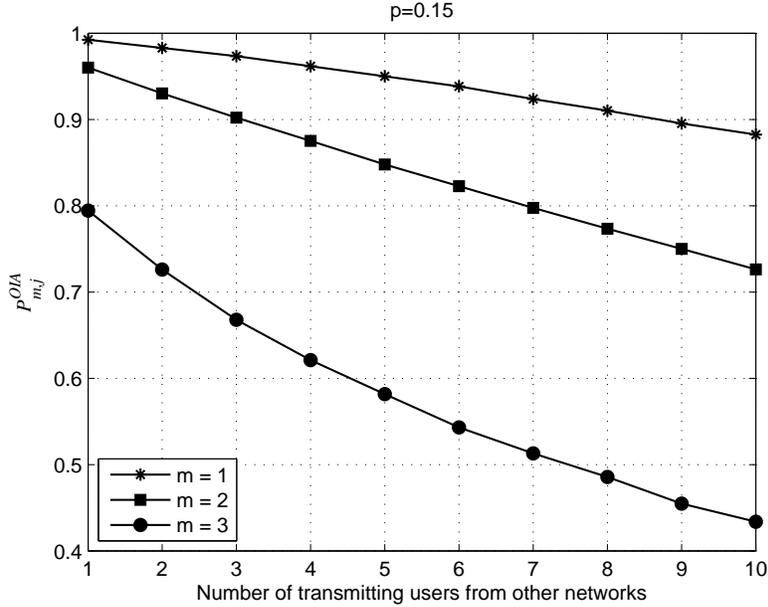}
    \caption{$P^{\rm OIA}_{m, j}$ vs. the number of simultaneously transmitting users from other networks ($K=3$, $L=3$, $M=3$, $p=0.15$).}
    \label{fig:Success_probability}
\end{figure}

We first consider the PHY layer performance of OIA for
the first RAN as a representative example. Fig.~\ref{fig:Success_probability} shows the
PHY-layer packet success probability of OIA, $P^{\rm OIA}_{m, j}$, by varying the number
of simultaneously transmitting (or active) users $(j)$ from other RANs when $p=0.15$ and
$S=3$. Three cases are considered where the numbers of concurrently transmitting users
$(m)$ in the first RAN are 1, 2, and 3. To investigate the steady state performance, the
CDF of LIF is obtained by collecting $10^5$ samples. We can observe that $P^{\rm
OIA}_{m, j}$ decreases with a larger $m$ or $j$. At the MAC layer, the number of
concurrently transmitting users, $m$ and $j$, vary slot by slot due to the random access
nature of the $p$-persistent protocol. Specifically, the first network sees local $m$
users' simultaneous transmissions with probability ${N\choose m} p^m (1-p)^{N-m}$ and
$j$ more users' simultaneous transmissions from the other RANs with probability ${(K-1)N
\choose j} p^j (1-p)^{(K-1)N}$, respectively. This phenomenon was implemented in the
simulations performed in this paper. If $m > S$ and $m+j>M$, a collision (or
transmission fail) happens. If $m \le S$, a smaller $P^{\rm OIA}_{m, j}$ happens with a
larger $m$ or $j$ as shown in Fig.~\ref{fig:Success_probability}. A larger $m$ indicates
a reduced DoF for the decoding at the users in the first RAN while a larger $j$
indicates a larger interference from the other RANs.

\begin{figure}
\centering
\includegraphics[width=4.5 in]{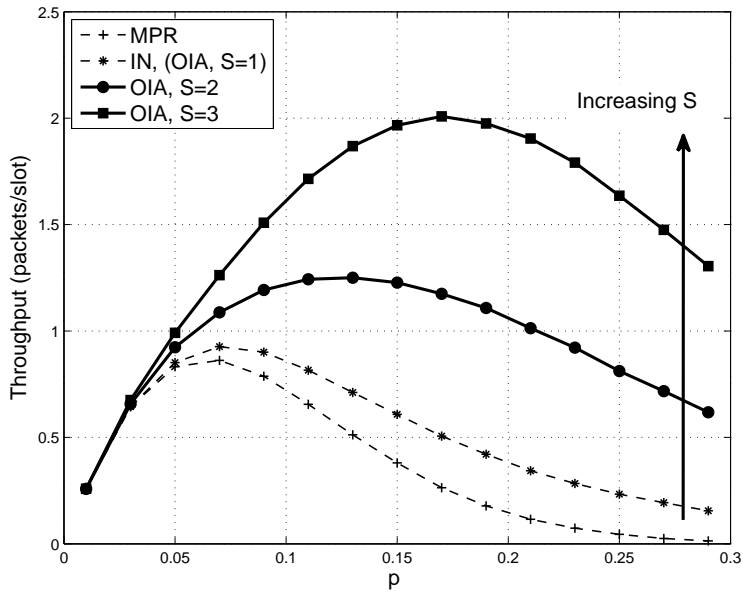}
    \caption{MAC layer throughput of the proposed OIA protocol according to the transmission probability~($p$).}
    \label{fig:Throughput_over_p}
\end{figure}

Fig.~\ref{fig:Throughput_over_p} compares the throughput of the
proposed OIA protocol with the conventional techniques including the
MPR and IN techniques for varying the transmission probability, $p$.
%The
%throughput performance of MPR, IN and OIA is shown in Fig.~\ref{fig:Throughput_over_p} where the transmission
%probability $p$ is varied form 0 to 0.3.
The proposed OIA protocol achieves much better throughput than the MPR
and IN techniques. Note that the IN technique in the figure is
identical to the proposed OIA protocol with $S=1$ because the condition
\eqref{eq:IN_limitation} is satisfied in this case and the OIA protocol
operates in the same way as the IN technique, as discussed in Section
IV. The IN technique achieves better throughput than the MPR technique
regardless of the transmission probability. The throughput increases as
the dimension of signal space in the OIA protocol~($S$) increases, and
thus we can conclude that a larger value of $S$ is preferable for the
proposed OIA protocol in this network scenario. For each scheme, there
exists the optimal transmission probability that maximizes the
throughput. We can observe that the optimal transmission probability of
the proposed OIA protocol increases as $S$ increases, which implies
that more aggressive transmission is preferable for larger $S$. For
example, the maximum throughput of the OIA protocol with $S=3$ is equal
to $2.01$ packets/slot, while that of the MPR technique is equal to
$0.86$ packets/slot. Hence, the OIA protocol yields $133\%$ throughput
improvement, compared to the MPR technique.

\begin{figure}
\centering
\includegraphics[width= 4.5in]{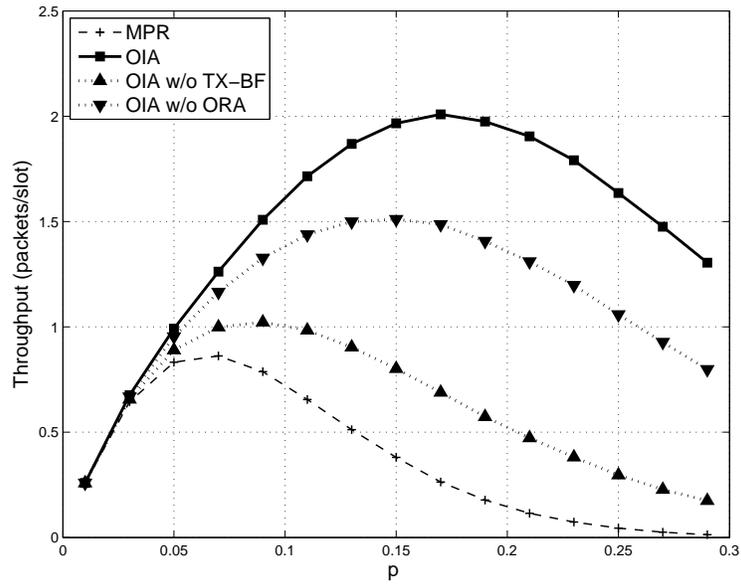}
    \caption{Throughput comparison among MPR, OIA protocol without transmit beamforming, OIA protocol without opportunistic random access, and the proposed OIA protocol according to the transmission probability when $S=3$.}
    \label{fig:Throughput_over_p_cross_N10_Effect}
\end{figure}

%{\em Effect of the SVD-based Beamforming and the opportunistic transmission:}
As explained in Section~\ref{sec:OIA}, the proposed OIA protocol
consists of transmit beamforming as a PHY layer technique and
interference-aware opportunistic random access as a MAC layer
technique. In order to analyze the contribution of each technical
component for overall throughput enhancement (also the overall effect
by joint design of two components),
%A fundamental question on the proposed OIA protocol is that how much each technical component affects the overall performance. Another question is why these two components are joinly required. To answer these questions,
we introduce two interference management protocols which exploit only one of the two
technical components of the proposed OIA protocol. We term the OIA protocol without
transmit beamforing at the PHY layer and the OIA protocol without opportunistic random
access in the MAC layer as `OIA w/o TX-BF' and `OIA w/o ORA', respectively.
% the effect the SVD-based beamforming and the opportunistic transmission on the throughput
%performance, we also perform simulations without the two functions which are individually named as `OIA-NoBeam' and
%`OIA-NoOpportunism' in Fig.~
Fig.~\ref{fig:Throughput_over_p_cross_N10_Effect} compares throughputs
of MPR, `OIA w/o TX-BF', `OIA w/o ORA', and the proposed OIA protocols.
In Fig.~ \ref{fig:Throughput_over_p_cross_N10_Effect}, $S$ is set to
$3$. The considered `OIA w/o TX-BF' and `OIA w/o ORA' protocols
outperform the MPR technique, while the effect of the SVD-based
transmit beamforming on the overall throughput is shown to be more
significant than that of the CDF-based opportunistic random access.
Compared to the OIA protocol without opportunistic random access, which
results in the maximum throughput of $1.02$ packets/slot, the proposed
OIA yields $97\%$ throughput enhancement. In addition, the proposed OIA
achieves $33\%$ thropughput improvement, compared to the OIA protocol
without transmit beamforming, which results in the maximum throughput
of $1.51$ packets/slot. It is found that the maximum throughput of the
proposed OIA protocol is achieved in the larger transmission
probability than the those of other schemes.

\begin{figure}
\centering
\includegraphics[width= 4.5in]{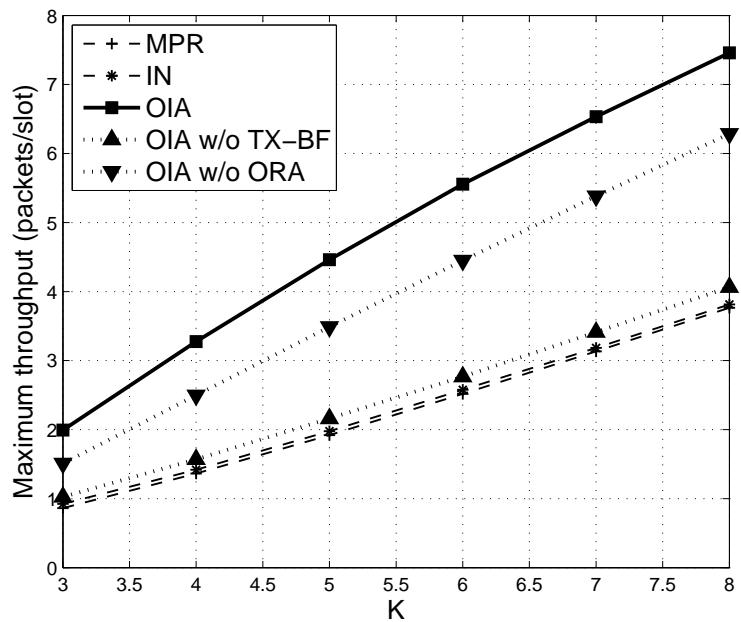}
    \caption{Maximum throughput of the considered protocols according to the number of overlapped RANs when $N=10$.}
    \label{fig:Throughput_over_K}
\end{figure}

Fig.~\ref{fig:Throughput_over_K} shows the maximum throughput of the several considered
protocols in this paper according to  the number of overlapped RANs~($K$) when there
exist $10$ users in each RANs~($N=10$). We also assume the numbers of transmit and
receive antennas as well as the dimension of signal space at APs are identical to the
number of overlapped RANs, i.e., $K=M=L=S$. The maximum throughput of each protocol is
evaluated by searching all possible transmission probabilities. From
Fig.~\ref{fig:Throughput_over_K}, we observe that the effect of the opportunistic random
access at the MAC layer on the maximum throughput is marginal. However, the effect of
the opportunistic random access at the MAC layer on the maximum throughput becomes
significant when it combines with the transmit beamforming at the PHY layer of the
proposed OIA protocol. Note that both the SVD-based transmit beamforming and the
CDF-based opportunistic random access play a role of reducing the interference among
overlapped RANs. From Fig.~\ref{fig:Throughput_over_K}, we can conclude that performance
gain from the opportunistic random access in the MAC layer can be magnified in
conjunction with the SVD-based transmit beamforming in the PHY layer. For example, the
proposed OIA protocol achieves $98.4\%$ throughput improvement, compared to the MPR
technique when $K=8$.

\section{Conclusion}\label{sec:Conclusion}
In this paper, we proposed a novel interference management protocol
called opportunistic interference alignment~(OIA) for overlapped random
access networks operating with slotted ALOHA, which intelligently
combines the interference alignment based transmit beamforming
technqiue at the PHY layer and the opportunistic random access
technique at the MAC layer. We also introduced a simple extension
method of the conventional techniques for interference-limited RANs:
multipacket reception and interference nulling. The proposed OIA
protocol is shown to significantly outperform the conventional schemes
in terms of MAC layer throughput. The proposed OIA protocol is expected
to be applied to next-generation wireless LANs such as IEEE 802.11 HEW
without significant modifications. We leave this issue for further
study.

%\appendix
%\section{Appendix}
%\subsection{Proof of Lemma~\ref{lem:converge}} \label{PF:converge}

%%%%%%%%%%%%%%%%%%%%%%%%%%%%%%%%%%%%%%%%%%%%%%%%%%%%%%%%%%%%%%%%%%%%%%%%%%%%%%%%%%%%%%%%%%%%%%%%%%%%%%%%%%%%%%%%%%%%%%%%%%%%%%%%%%%%%%%%%%%%%%%%%%%%%

\end{document}